\documentclass[twocolumn,preprintnumbers,amsmath,amssymb]{revtex4-2}
\usepackage{graphicx}  
\usepackage{sidecap}
\begin{document}
%
\title{Effect of electron thermal conductivity on resonant plasmonic detection  in the metal/black-AsP/graphene FET terahertz  hot-electron bolometers 
}
\author{V.~Ryzhii$^{1}$, C.~Tang$^1$, T.~Otsuji$^1$, M.~Ryzhii$^{2}$,
V.~Mitin$^3$, and M.~S.~Shur$^4$}
\address{
$^1$Research Institute of Electrical Communication,~Tohoku University,~Sendai~ 980-8577, 
Japan\\
$^2$Department of Computer Science and Engineering, University of Aizu, Aizu-Wakamatsu 965-8580, Japan\\
$^3$Department of Electrical Engineering, University at Buffalo, SUNY, Buffalo, New York 14260 USA\\
$^4$Department of Electrical,~Computer,~and~Systems~Engineering, Rensselaer Polytechnic Institute,~Troy,~New York~12180,~USA
}

\begin{abstract} 
\normalsize We analyze the two-dimensional electron gas (2DEG) heating by the incident terahertz (THz) radiation
in the field-effect transistor (FET) structures with the graphene channels (GCs) and the black-phosphorus  and black-arsenic  gate barrier layers (BLs). Such GC-FETs can operate as
bolometric THz detectors using the thermionic emission of the hot electrons from
the GC via the BL into the gate. Due to the excitation of plasmonic oscillations in the GC by the THz signals, the GC-FET detector response can be pronouncedly resonant, leading to elevated values of the detector responsivity.
The  lateral thermal conductivity of the 2DEG can markedly affect  the GC-FET responsivity, in particular, its spectral characteristics. This effect should be considered for the optimization of the GC-FET detectors.
\end{abstract} 
\maketitle

\section{Introduction}
The field-effect transistor (FET) structures with the graphene channel (GC) and the gate barrier layer (BL) made of
 the black-Phosphorus (b-P), black-Arsenic (b-As), and the compounds of these materials (b-As$_{1-x}$P$_x$) have shown promise for various device applications. This is because of the unique electron properties of GCs~\cite{1} and  the b-As$_{1-x}$P$_x$ layers~\cite{2,3} (predicted and demonstrated a long time ago~\cite{4,5}).
 Changing the As mole fraction $x$ and the BL thickness $W$~\cite{6,7,8} allows to achieve a desirable band alignment and the height of the barriers between GC and the gate, therefore controlling the thermionic activation energy.
The latter opens up additional prospects for the FET's optimization and use for novel devices, in particular, photodetectors\cite{9,10,11,12,13,14,15,16,17,18,19,20,21}. 
 Recently~\cite{22}, we proposed and evaluated the hot-electron bolometric detector of terahertz radiation (THz) based on such FET structures.
 Our analysis confirmed the controllability of the electron activation energy by doping and the gate bias voltage and 
 predicted high values of the detector responsivity  associated with the effective heating of the two-dimensional electron gas (2DEG) in the GC by the signal electric field produced by the impinging THz radiation, especially  when the plasmonic oscillations are resonantly excited.

 In this paper, we generalize the  device model of GC-FET detectors with b-P and b-As
 BLs by accounting for the following effects:\\
 (1) the electron thermal conductance along the GC and\\ 
 (2) the electron cooling due to the thermionic emission from the GC into the gate (the Peltier cooling). 
 The role of the electron thermal conductivity  can be crucial since it is   fairly high (see, for example,~\cite{23,24}).
 Thermionic cooling
 can become essential at elevated gate bias voltages when the effective
 height of the energy barrier for electrons (holes) decreases.
 These effects can influence the detector's performance and should be taken into account for its optimization.

\begin{figure*}[t]\centering
\includegraphics[width=13.0cm]{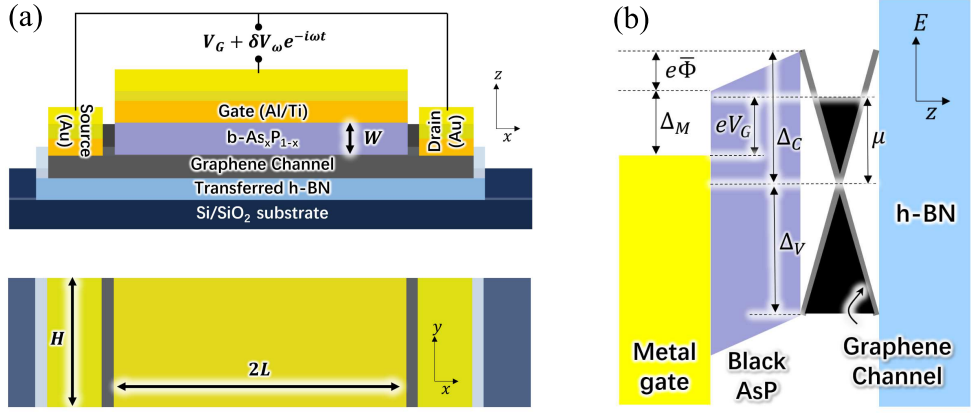}
\caption{
(a) Cross-section and top view of the GC-FET detector structure with the Al or Ti gate,
b-As$_{1-x}B_{x}$ BL , and the GC on the h-BN substrate (b) its band diagram under the applied  dc bias and ac signal voltages
  $V_G + \delta V_{\omega}\exp(-i\omega t)$ between the gate the GC side contacts~\cite{22}. 
} 
\label{F1}
\end{figure*}

\section{Device structure}
\vspace{-0.3cm}
Figure 1 shows the GC-FET structure cross-section and top view and its band diagram when the gate bias voltage $V_G  \gtrsim 0$. For the definiteness, the materials for the BL and the metal gate, and the GC doping are chosen to provide the flat band in the BL at $V_G =0$.  This band diagram configuration corresponds to the Ohmic side contacts (the FET source and drain).
In particular, we focus on the GC-FETs with the b-P and b-As BLs and Al and Ti gates
(the Al/b-P/GC and Ti/b-As/GC devices).
The GC and the gate length (the spacing between the side contacts, their width, and the thickness of the BL are set to be $2L$, $H$, and $W$, respectively 
(with $2L, H \gg W$).
We focus on the GC-FETs with the GC doped by donors. We assume that the activation energy for
the electrons in the GC $\Delta_C - \mu_D = \Delta_M > \Delta_V$, where $\Delta_C$ and  $\Delta_V$
are the band offsets between the BL conduction and valence bands and the Dirac point in the GC,
and $\Delta_M $ is the difference between the BL and the GC work functions.
Here $\mu_D$ is the electron Fermi energy in the GC at $V_G=0$, which is determined by the donor density $\Sigma_D$. 
The latter inequality implies that the thermionic electron gate current  dominates over the hole current. The GC-FET devices in which the 
above condition is not valid  can be considered analogously. 

The voltage between
the GC side contacts and the highly conducting metal  gate comprises, 
apart from the bias voltage $V_G$,  the ac component  $\delta V_{\omega}\exp(-i\omega\,t)$, where $\delta V_{\omega}$ and $\omega$ are the signal amplitude and the frequency.
This is  the signal generated in an antenna by incident THz radiation. The THz signal could excite  the plasmonic oscillation modes with the symmetric spatial distribution of the ac potential in the GC.

\section{2DEG heating}

The variation of the thermionic gate current, $\delta j$,  associated with the incoming THz signal is given by

\begin{eqnarray}\label{eq1}
\delta j = j^{max}\frac{\delta T}{T}\biggl(\frac{\Delta_C-\mu}{T}\biggr)\exp\biggl(-\frac{\Delta_C-\mu}{T}\biggr),
\end{eqnarray} 
where $j^{max}$ is the maximal value of the current density from the GC,
which we estimate as $j^{\max}= e\Sigma/\tau_{\bot}$ with $\Sigma$ and $\tau_{\bot}$ being the 2DEG density in the GC  and the escape time, respectively, of the electrons with the energy exceeding the barrier height.
Both $\Sigma$ and $\tau_{\bot}$ depend on doping and the gate bias voltage.

The quantity $\delta j$ comprises 
the ac component $\delta j_{\omega}\exp(-i\omega t)$ 
and the rectified dc component $<\delta j_{\omega}>$ (as well as the second and higher harmonics). 
The latter is
the variation of the current density averaged over the period of the THz signal $2\pi/\omega$. According to Eq.~(1), the rectified component can be presented as

\begin{eqnarray}\label{eq2}
<\delta j_{\omega}> = j^{max}{\mathcal F}\frac{<\delta T_{\omega}>}{T},
\end{eqnarray} 
where 
\begin{eqnarray}\label{eq3}
{\mathcal F} = 
\frac{\Delta_C-\mu}{T}\exp\biggl(-\frac{\Delta_C-\mu}{T}\biggr).
\end{eqnarray} 

The dependence of the factor ${\mathcal F}(V_G)$ on the gate voltage  is associated with 
the voltage dependence of  the electron Fermi energy  (see below).

The variation of the local averaged 2DEG temperature, $ <\delta T_{\omega}>$ is governed by the following 
electron heat transport equation:

\begin{eqnarray}\label{eq4}
-h\frac{d^2 <\delta T_{\omega}>}{d x^2}+ 
\frac{ <\delta T_{\omega}>}{\tau_{\varepsilon}} 
= Q_{\omega}^J - Q_{\omega}^B.
\end{eqnarray} 
Here 
$h \simeq v_W^2/2\nu$ is the  electron thermal conductivity in the GC per electron
(this  corresponds to the Wiedemann-Franz relation),   $\tau_{\varepsilon}$ is the electron energy relaxation time in the GC, $v_W \simeq 10^8$~cm/s is the characteristic electron velocity in GCs, and $\nu$ is the electron scattering frequency in the GC. The second term in the left side of Eq.~(4) is associated with the 2DEG energy transfer to the phonon system (particularly to  optical phonons), while the terms in the right site, $Q_J$ and $Q_B$,
are the Joule power received by the  2DEG in the GC and
the power transferred by the electrons emitted from the 2DEG over the BL averaged over time (see, for example,~\cite{25}).
 The Joule contribution to the 2DEG heating is given by

\begin{eqnarray}\label{eq5}
Q_{\omega}^J
= \frac{{\rm Re} \sigma_{\omega}}{\Sigma}<|\delta E_{\omega}|^2>.
\end{eqnarray} 
Here,
accounting for the excitation of plasmonic oscillations in the gated 2DEG (see, for example, ~\cite{26,27,28,29,30,31,32,33,34,35,36} and the references therein) by
the THz signal received by an antenna, one can obtain for the ac potential in the GC ~\cite{22}

\begin{eqnarray}\label{eq6}
\delta \varphi_{\omega}
= \delta V_{\omega}\frac{\cos (\gamma_{\omega}x/L)}{\cos (\gamma_{\omega})}.
\end{eqnarray} 
Hence for the square of the signal electric field amplitude 
 we arrive at

\begin{eqnarray}\label{eq7}
<|\delta E_{\omega}|^2> = \frac{1}{2}\biggl(\frac{\delta V_{\omega}}{L}\biggr)^2\biggl|\frac{\gamma_{\omega}\sin(\gamma_{\omega}x/L)}{\cos \gamma_{\omega}}\biggr|^2,
\end{eqnarray} 
where Re~$\sigma_{\omega} = \sigma_0\nu^2/(\nu^2+\omega^2)$, $\sigma_0 = e^2\mu/\pi\hbar^2\nu$ and $\nu$ are the Drude dc conductivity and the electron scattering frequency in the 2DEG,  $\gamma_{\omega} =\pi\sqrt{\omega(\omega+i\nu)}/2\Omega_P$ and
$\Omega_P=(\pi\,e/\hbar\,L)\sqrt{\mu\,W/\kappa}$ are the effective wavenumber and the plasmonic frequency (corresponding to the symmetric conditions at the contacts), respectively, with
$\kappa$ and $W$ being the dielectric constant of the BL  and its thickness.
The quantity $Q_{\omega}^B$, which characterizes the cooling of the 2DEG,
is presented as 

\begin{eqnarray}\label{eq8}
Q_{\omega}^B= \frac{<\delta j_{\omega}>}{e\Sigma}\Delta_C = \frac{{\mathcal F}\Delta_C}{\tau_{\bot}}\frac{<\delta T_{\omega}>}{T}.
\end{eqnarray} 

\section{Rectified current}

\begin{figure*}\centering
\includegraphics[width=13.5cm]{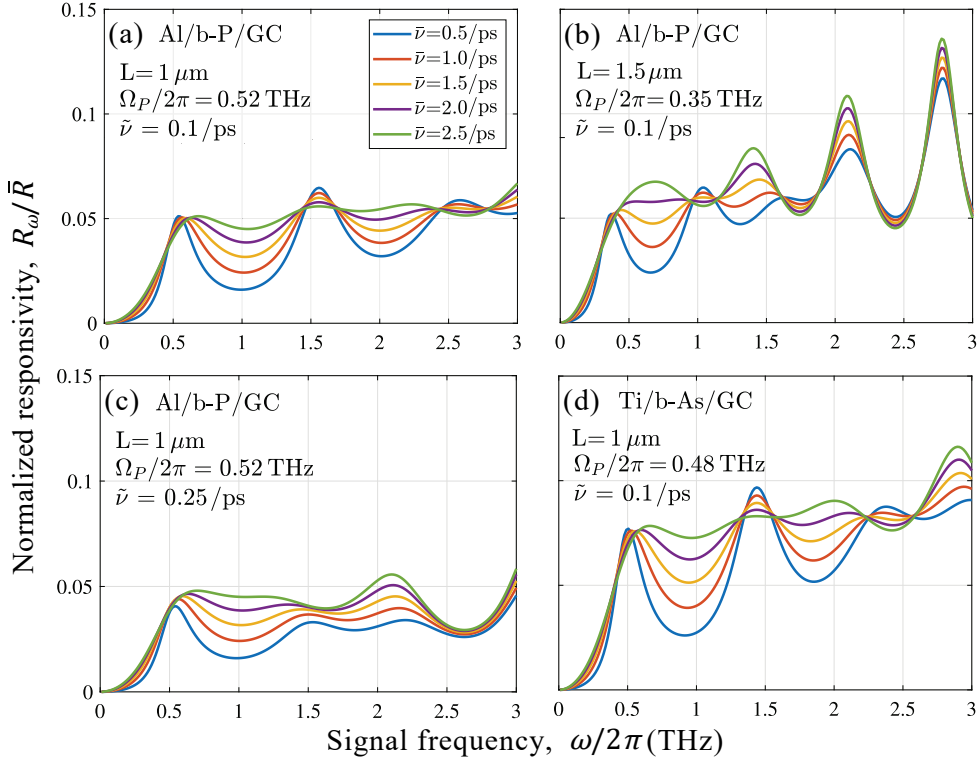}
\caption{Normalized responsivity $R_{\omega}/{\overline R}$ versus signal frequency $\omega/2\pi$ of the GC-FETs (a) with the  Al/b-P/GC structure and 
 the GC half-length $L = 1~\mu$m  ($\Omega_P/2\pi = 0.52$~THz) and ${\tilde \nu} = 0.1$~ps$^{-1}$, (b)  $L = 1.5~\mu$m  ($\Omega_P/2\pi = 0.35$~THz) and ${\tilde \nu} = 0.1$~ps$^{-1}$, (c) $L = 1~\mu$m  ($\Omega_P/2\pi = 0.52$~THz) and ${\tilde \nu} = 0.25$~ps$^{-1}$, and (d) with Ti/b-As/GC structure and $L = 1~\mu$m  ($\Omega_P/2\pi = 0.48$~THz) and ${\tilde \nu} = 0.1$~ps$^{-1}$.
} 
\label{F2}
\end{figure*}

Assuming sufficiently low thermal resistance of the highly conducting side contacts, we set
the following boundary conditions for Eq.~(4):

\begin{eqnarray}\label{eq9}
 <\delta T_{\omega}>|_{x = \pm L} =0,\qquad <\delta j_{\omega}>|_{x = \pm L} =0.
\end{eqnarray} 

Solving Eq.~(4) with the boundary conditions given by Eq.~(9) and invoking Eq.~(2),
we arrive at the following spatial distribution of the rectified current density:

\begin{eqnarray}\label{eq10}
- {\mathcal L}^2\frac{d^2 <\delta j_{\omega}>}{d x^2}+ 
 <\delta j_{\omega}>\nonumber\\
=\frac{e{\rm Re}\sigma_{\omega}}{2T}\frac{\tau_{\varepsilon}{\mathcal F}/\tau_{\bot}}
{(1 + \tau_{\varepsilon} {\mathcal F}\Delta_C/
\tau_{\bot}T)}
 \biggl|\frac{\gamma_{\omega}\sin(\gamma_{\omega}x/L)}{\cos \gamma_{\omega}}\biggr|^2\biggl(\frac{\delta V_{\omega}}{L}\biggr)^2,
\end{eqnarray} 
where
${\mathcal L} = \sqrt{h\tau_{\varepsilon}/(1+\tau_{\varepsilon} {\mathcal F}\Delta_C/\tau_{\bot}T)}$
is the "thermal" length.
The spatial dependence in the right-hand side of Eq.~(10) reflects the nonuniformity of the 2DEG heating associated with the nonuniform distribution of the ac plasmonic electric field along the GC.

The rectified current density $<\delta j_{\omega}>$ is an oscillatory function of the signal frequency. It exhibits the resonant peaks at $\omega \simeq (2n-1)\Omega_P$, where $n= 1,\,2,\,3,...$ is the index of the plasmonic resonance, provided that $\nu \ll \Omega_P$.
The latter inequality is assumed in the following.

\begin{widetext}
Considering the condition  $\nu \ll \Omega_P$, Eq.~(10)
can be somewhat simplified and presented as

\begin{eqnarray}\label{eq11}
- {\mathcal L}^2\frac{d^2 <\delta j_{\omega}>}{d x^2}+ 
 <\delta j_{\omega}>
\simeq 
\frac{2e\sigma_{0}}{T}\frac{\tau_{\varepsilon}{\mathcal F}/\tau_{\bot}}
{(1 + \tau_{\varepsilon} {\mathcal F}\Delta_C/
\tau_{\bot}T)}\biggl(\frac{\delta V_{\omega}}{L}\biggr)^2
\frac{\sin^2[\pi\omega/2\Omega_P)(x/L)]}
{\sin^2(\pi\omega/2\Omega_P)+ (4\Omega_P/\pi\nu)^2\cos(\pi\omega/2\Omega_P)^2]}.
\end{eqnarray} 

Introducing the parameter $\ae = {\mathcal L}/L$ 
and solving Eq.~(10) with boundary conditions~(9), we obtain

\begin{eqnarray}\label{eq12}
 <\delta j_{\omega}> = 
\frac{e\sigma_{0}}{T}\frac{\tau_{\varepsilon}{\mathcal F}/\tau_{\bot}}
{(1 + \tau_{\varepsilon} {\mathcal F}\Delta_C/
\tau_{\bot}T)}\biggl(\frac{\delta V_{\omega}}{L}\biggr)^2
\Biggl\{\frac{1 - \displaystyle\frac{\cos(\pi\,\omega\,x/\Omega_PL)}{1+(\pi\ae\,\omega/\Omega_P)^2}
- \displaystyle\biggl[1-\frac{\cos(\pi\,\omega/\Omega_P)}{1 + (\pi\ae\,\omega/\Omega_P)^2}\biggr]
\frac{\cosh(x/L\ae)}{\cosh(1/\ae)}
}{\sin^2(\pi\,\omega/2\Omega_P)+ (4\Omega_P/\pi\nu)^2\cos^2(\pi\,\omega/2\Omega_P)}\Biggr\}.
\end{eqnarray}

Integrating $<\delta j_{\omega}>$, given by Eq.~(12), over the GC  plane, for the net rectified current $<\delta J_{\omega}> = H\int_L^Ldx<\delta j_{\omega}>$ at the plasmonic resonances,
we arrive at the following expression:

\begin{eqnarray}\label{eq13}
<\delta J_{\omega}> =
\frac{2HLe\sigma_{0}}{T}\frac{\tau_{\varepsilon}{\mathcal F}/\tau_{\bot}}
{(1 + \tau_{\varepsilon} {\mathcal F}\Delta_C/
\tau_{\bot}T)}\biggl(\frac{\delta V_{\omega}}{L}\biggr)^2 \biggl[\frac{\Pi_{\omega}(\ae)}
{[\sin^2(\pi\,\omega/2\Omega_P)+ (4\Omega_P/\pi\nu)^2\cos^2(\pi\,\omega/2\Omega_P)]}.
\end{eqnarray} 
Here \begin{eqnarray}\label{eq14}
\Pi_{\omega}= 
1 - \displaystyle\frac{\sin(\pi\,\omega/\Omega_P)}{(\pi\,\omega/\Omega_P)[1+(\pi\ae\,\omega/\Omega_P)^2} -
 \biggl[1-\displaystyle\frac{\cos(\pi\,\omega/\Omega_P)}{1+(\pi\ae\,\omega/\Omega_P)^2}\biggr]\ae\tanh\biggl(\frac{1}{\ae}\biggr).
\end{eqnarray}

The factor $\Pi_{\omega}(\ae)$ characterizes the effect of the electron thermal conductivity on
the rectified current and the signal frequency dependence of the latter.
\end{widetext}

\section{GC-FET Detector responsivity.}

The current responsivity (in the units A/W) of the GC-FET detector $ R_{\omega}$ is proportional to the 
net rectified current $\overline {<\delta J_{\omega}>}$, which is given by Eqs.~(13) and (14),
divided  the incident THz radiation power $P_{\omega}$ collected by an antenna.
Considering that the relation between  the ac voltage generated
at the side contacts $\delta V_{\omega}$ and the power $P_{\omega}$~\cite{36}  is given by $\delta V_{\omega}^2 = 16\pi^2P_{\omega}/c$ (with
$c$ being the speed of light in vacuum), we obtain
 the GC-FET bolometric detector responsivity  $R_{\omega} =\overline {<\delta J_{\omega}>}/P_{\omega}$

\begin{eqnarray}\label{eq15}
R_{\omega} = {\overline R} \frac{(\tau_{\varepsilon}{\mathcal F}\Delta_C/\tau_{\bot}T)}
{(1 + \tau_{\varepsilon} {\mathcal F}\Delta_C/
\tau_{\bot}T)}\biggl(\frac{v_W}{L}\biggr)
\frac{\Pi_{\omega}(\ae) }{[{\overline \nu}+ {\tilde \nu}\,(\omega/\Omega_P)^2]}.
\end{eqnarray}
Here

\begin{eqnarray}\label{eq16}
 {\overline R} = 
 \frac{32\pi}{137}\frac{e}{\Delta_C}
 \frac{\mu}{\hbar}\frac{H}{v_W},
\end{eqnarray}
where we assumed that  the antenna gain $g = 2$~\cite{37}  and took into account that the fine structure parameter $\alpha =e^2/c\hbar \simeq 1/137$ .
To account for the effect of the 2DEG viscosity on the plasma oscillations damping in Eq.~(15), we set~\cite{38} $\nu = {\overline \nu} + {\tilde \nu}\,(\omega/\Omega_P)^2$,
where
${\overline \nu}$ and ${\tilde \nu}$ are related to the electron interaction with impurities and acoustic phonons and  the electron viscosity of the  2DEG in the GC, respectively.

 Using Eq.~(16), for  room temperature 
for $\mu = 120 - 140$~meV  and $H= 5 -10~\mu$m,
we obtain the estimate Eq.~(16)  yields  ${\overline R} 
\simeq (3 - 6) \times 10^3$~A/W. \\

\section{Spectral characteristics}
\vspace{-3mm}
\begin{figure*}[t]\centering
\includegraphics[width=13.5cm]{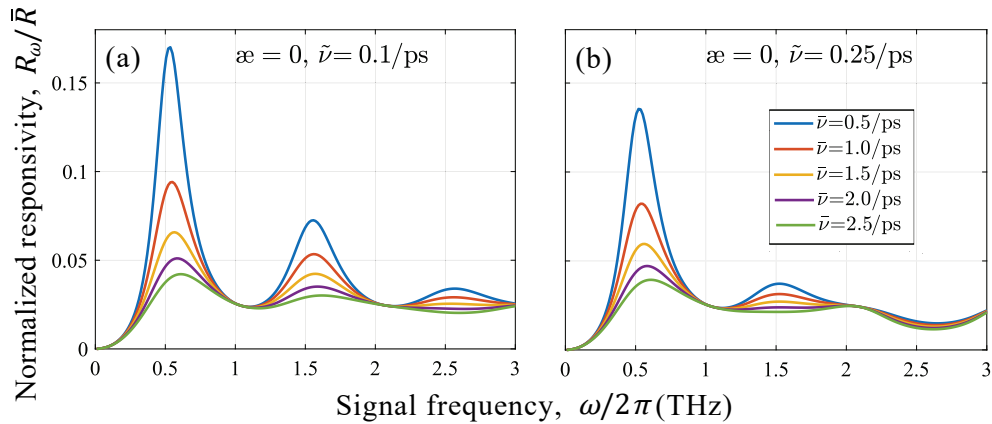}
\caption{The same as for Fig.~2(a), but for weak electron thermal transport   (in the limit $\ae = 0$)
.}
\label{F3}
\end{figure*}

Below we calculate the room temperature  GC-FET detector responsivity as a function of the signal frequency 
for different plasmonic and collision frequencies and 
 bias voltages using Eqs.~(15) and (16) and invoking Eq.~(14). 
As examples, we consider the   GC-FETs with 
the band parameters listed in Table~I. Other parameters are indicated in the figure captures. 

                                                                                                                                                                                                                                                                                                                                                                                                                                                                                                                                                                                                                                                                                                                                                                                                                                                                                                                                                                                                                                                                                                                                                                                                                                                                                                                                                                                                                                                                                                                                                                                                                                                                                                                                                                                                                                                                                                                                                                                                                                                                                                                                                                                                                                                                                                                                                                                                                                                                                                                                                                                                                                                                                                                                                                                                                                                                                                                                                                                                                                                                                                                                                                                                                                                                                                                                                                                                                                                                                                                                                                                                                                                                                                                                                                                                                                                                                                                                                                                                                                                                                                                                                                                                                                                                                                                                                                                                                                                                                                                                                                                                                                                                                                                                                                                                                                                                                                                                                                                                                                                                                                                                                                                                                                                                                                                                                                                                                                                                                                                                                                                                                                                                                                                                                                                                                                                                                                                                                                                                                                                                                                                                                                                                                                                                                                                                                                                                                                                                                                                                                                                                                                                                                                                                                                                                                                                                                                                                                                                                                                                                                                                                                                                                                                                                                                                                                                                                                                                                                                                                                                                                                                                                                                                                                                                                                                                                                                                                                                                                                                                                                                                                                                                                                                      \begin{table}[]
\centering
\vspace{2 mm}
\begin{tabular}{|r|c|c|c|c|c}
\hline
Structure\,&   $\Delta_M$(meV)& $\Delta_V$(meV)& $\Delta_C$(meV)& $\mu_D$~(meV)\\ 
\hline
Al/b-P/GC\,&	85&	125& 225& 140 \\ 
\hline
Ti/b-As/GC& 70&	90&	  190& 120\\
\hline
\end{tabular}
\caption{\label{table} Parameters of the GC-FET structures under consideration~\cite{39,40,41}.} 
\end{table}

 Figure~2 shows the spectral dependences of the normalized responsivity of the GC-FET bolometric detectors with the  Al/b-P/GC and Ti/b-As/GC structures
 calculated for different values of the  frequencies ${\overline \nu}$ and ${\tilde \nu}$ (and, hence, different electron thermal lengths) and different GC lengths at $T = 25$~meV and $V_G \simeq 0$.  The latter implies that ${\mathcal F} \simeq {\mathcal F}_0 = (\Delta_M/T)\exp(-\Delta_M/T))$.
 The structural parameter ranges are discussed below in Appendix A. 

The frequency dependences of the responsivity demonstrated in Figs.~2 and 3 exhibit
 the pronounced
peaks at the frequencies close to  the plasmonic resonant frequencies $\omega = (2n-1)\Omega_P$. 
As follows from the comparison of Figs.~2(a) and 2(c),  an increase in the electron viscosity markedly smears  the higher plasmonic resonances.

One can see from Fig. 2 that the responsivity $R_{\omega}$  can  be markedly smaller than its characteristic value ${\overline R}$, even at the resonant maxima (about one order of magnitude  smaller as follows, for example,  from Eq.~(17) and the consequent estimate).

\section{Role of electron  thermal conductivity }

An increase in the collision frequency ${\overline \nu}$ leads, as might be expected, to
a smearing of the resonant peaks. However, the average values of the responsivity
show a moderate increase with increasing ${\overline \nu}$. This can be interpreted 
by a smaller  electron thermal conductivity $h$ in the GCs with a larger ${\overline \nu}$. As a result, increasing collision frequency ${\overline \nu}$ gives rise to  a smaller Joule power but  a weaker transfer of the 2DEG heat to the side contacts.
The comparison of Figs. 2(a) and 2(b)  indicates that
in the devices with a longer GC, the peaks corresponding to  higher resonances
can be much more pronounced. This is because an increase in $L$ leads to a weakening
of the electron thermal transfer (the parameter $\ae \propto L^{-1}$). 

To confirm the significance of the electron thermal transfer in the GC-FETs under consideration, we calculated the frequency dependences  of the responsivity disregarding
the electron thermal conductivity, i.e., neglecting the electron heat transfer to
the side contacts.  
Figure~3 shows the responsivity of the GC-FET with the parameters used for Figs.~2(a)
and 2(c) 
calculated for the  hypothetical case when the electron thermal conductivity is insignificant. One can see that
in the case of $\ae \rightarrow 0$, the resonant peaks are higher and their smearing is weaker. The plots of Fig.~3 are similar to those in~\cite{22} (despite a marked difference in the electron viscosity strength).
Thus, the electron thermal transfer along the GC to the side contacts pronouncedly affects the responsivity spectral characteristics.  
The comparison of the plots in Fig.~3(a) and 3(b) indicates that the electron viscosity
naturally weakens the plasmonic resonances.

\section{Comments}

An increase in the gate voltage $V_G$ results in the rise of $R_{\omega}$ associated with a larger ${\mathcal F}$ factor in some range of  larger voltages (see Appendices A and B). Considering Eq.~(3) and accounting for Eq.~(B3), for not too large $V_G$, we obtain

\begin{eqnarray}\label{eq17}
{\mathcal F} \simeq \frac{\Delta_M - \mu_0eV_G/(\mu_0 + \mu_D)}{T}
\nonumber\\
\times\exp\biggl[-\frac{\Delta_M - \mu_0eV_G/(\mu_0 + \mu_D)}{T}\biggr].
\end{eqnarray}
Here $\mu_0 = (\hbar^2v_W^2\kappa)/8e^2W$ (see Appendices A and B).
As follows from Eq.~(17), the sensitivity of the factor ${\mathcal F}$ to the gate voltage $V_G$  is weakened by a relatively small parameter $\mu_0/(\mu_0 + \mu_D)$.

In the expression for the signal current density given by Eq.~(2), we disregarded the rectified component associated with the modulation of the activation energy by the plasmonic oscillations potential. 
The thermionic current density as a function of the net voltage drop is determined by by the following factor: $\displaystyle\frac{\mu_0}{(\mu_0+\mu_D)}
\frac{e(V_G+ \delta\varphi_{\omega})}{T}$, where $\delta\varphi_{\omega}$
is given by Eq.~(6).
At low gate voltage, the pertinent ac current density component is equal to
 
\begin{eqnarray}\label{eq18}
\langle \delta {\tilde j_{\omega}}\rangle \simeq \frac{j^{max}}{2}
\exp\biggl(- \frac{\Delta_M}{T} \biggr) \biggl(\frac{\mu_0}{\mu_0+\mu_D}\biggr)^2
\frac{\langle \delta \varphi_{\omega}^2\rangle}{T^2}\nonumber\\
\simeq  \frac{j^{max}}{4}\exp\biggl(- \frac{\Delta_M}{T}\biggr)\biggl(\frac{\mu_0}{\mu_0+\mu_D}\biggr)^2\biggl|\frac{\cos(\gamma_{\omega}x/L)}{\cos \gamma_{\omega}}\biggr|^2 \frac{(\delta V_{\omega})^2}{T^2}.\,
\end{eqnarray}
Comparing the thermionic rectified current density given by Eq.~(2) and the consequent formulas
 with $\langle \delta {\tilde j_{\omega}}\rangle$,
described by Eq.~(18), at the pronounced plasmonic resonances ($\Omega_P \gg \nu$), we obtain 

\begin{eqnarray}\label{eq19}
\frac{\langle \delta  j_{\omega}\rangle}
{\langle \delta {\tilde j_{\omega}}\rangle} 
\simeq \biggl(\frac{\kappa}{\pi}\biggr) 
\frac{\tau_{\varepsilon}\nu\Delta_M}{e^2W\Sigma}
\biggl(\frac{\mu_0+\mu_D}{\mu_0}\biggr)^2\nonumber\\
\simeq 
8(\tau_{\varepsilon}\nu)\biggl
(\frac{\mu_0 + \mu_D}{\mu_D}\biggr)^2\biggl(\frac{\Delta_M}{\mu_0}\biggr)\simeq 8(\tau_{\varepsilon}\nu)\biggl(
\frac{\Delta_M}{\mu_0}\biggr).
\end{eqnarray}
 Since the factors $\tau_{\varepsilon}\nu$ and $\Delta_M/\mu_0$ in the right-hand side of Eq.~(19) are much larger than unity, $\langle \delta  j_{\omega}\rangle/\langle \delta {\tilde j_{\omega}}\rangle \gg 1$.


Apart from the detector responsivity $R_{\omega}$ analyzed above,
the responsivity, $R_{\omega}^{area}$, normalized by the detector area $2lH$ can be used. For $R_{\omega}^{area}$ one can obtain $R_{\omega}^{area} = R_{\omega}(S_{\omega}/2LH)$, where $S_{\omega} = \lambda_{\omega}^2g/4\pi$ is the antenna aperture~\cite{36} and $\lambda_{\omega} 2\pi\,c/\omega$ is the wavelength of the incident radiation in vacuum.  This implies
that $R_{\omega}^{area} \gg R_{\omega}$, at least in the THz range, and
$R_{\omega}^{area} \propto R_{\omega}/\omega^2$, hence $R_{\omega}^{area}$ and $R_{\omega}$ exhibit fairly different spectral characteristics.

Using Eqs.~(14) and (18), we find that $R_{\omega}$ reaches a maximum at
$V_G = V_G^{max} = (\Delta_M - T)(\mu_0 + \mu_D)/e\mu_0$. Setting, for example,
$\Delta_M = 85$~meV,  $\mu_D = 140$~meV, and $\mu_0 =(14 - 21)$~meV, we obtain
$V_G^{\max} \simeq  (460 - 660)$~meV. However, it is impractical to use such large gate voltages because this can lead to very high DC density [given by Eq.~(B4)], at which the spatial distribution of the steady-state temperature of the 2DEG can be substantially  nonuniform, particularly, when the current-crowding effect might become crucial.
Moreover, at the voltages in question, the electron tunneling between the GC and the gate can enable an elevated tunneling current (see Appendix C). Although the tunneling current does not markedly affect the bolometric response (due to its weak dependence on the 2DEG effective temperature), it together with the elevated thermionic current can substantially decrease the GC-FET detector dark-current limited detectivity. 

This implies that the maximal GC-FET detector detectivity can be achieved at small $V_G$ when the dark current density ${\overline j}$ is fairly small, being primarily limited not by  the dark current but by the Johnson-Nyquist  noise  (see, for example,~\cite{42}).

\section*{Conclusions}

We evaluate the effect of the electron thermal conductivity on the characteristics
of the 
GC-FET bolometric
THz detectors with the Al/b-P/GC and Ti/b-As/GC. The operation of such devices is enabled by the heating of   
the 2DEG  in the GC by the THz radiation resulting in an increase in the thermionic current from the GC via the b-P or b-As BLs into a metal gate.
We demonstrate that the electron thermal conductivity in the GC can markedly influence the 2DEG energy balance pronouncedly affecting the value and the  spectral characteristics of the GC-FET detectors, which should be accounted for in the device optimization.   

\section*{Author's contributions}
All authors contributed equally to this work.

\section*{Acknowledgments}
The Japan Society for Promotion of Science (KAKENHI
Grants $\#$ 21H04546 and $\#$ 20K20349), Japan;
RIEC Nation-Wide Collaborative Research Project $\#$
R04/A10; the US Office of Scientific, Research Contract
N00001435, (Project Monitor Dr. Ken Goretta).

\section*{Data availability}
The data that support the findings of this study are available within the article.

\subsection*{Appendix A. Device parameters} 
 \setcounter{equation}{0}
\renewcommand{\theequation} {A\arabic{equation}}
We assume that
the electron energy relaxation time $\tau_{\varepsilon}$ at  room temperature is determined primarily by the GC optical phonons~\cite{43,44,45,46,47} and the interface optical phonons~\cite{48,49}. The analysis of these references, considering that the time of the spontaneous optical phonon emission in GC $\tau_0 \lesssim 1$~ps and
accounting for the smallness of the optical phonon number at room temperature,  leads to the estimate $\tau_{\varepsilon} \simeq 10 - 20$~ps. 
The escape of an electron with an energy exceeding $\Delta_C$ from 2DEG into the gate via the BL is possible after its scattering  on acoustic phonons
and impurities with a substantial variation of the electron  momenta (with the electron turning  almost perpendicular to the GC plane). 
Assuming that the electron  momentum relaxation time associated with the acoustic phonon scattering at room temperature $\tau_{ac} \sim 1$~ps~\cite{50,51,52},
one can conclude that $\tau_{\bot} >  \tau_{ac}$ or even  $\tau_{\bot} \gg \tau_{ac}$.
Considering this, in our estimates and calculations below, we set $\tau_{\bot} \sim 10$~ps.

The  voltage factor $ (\tau_{\varepsilon} {\mathcal F}/\tau_{\bot})/[1+(\tau_{\varepsilon} {\mathcal F}\Delta_C/\tau_{\bot}T)]$ in Eqs.~(9) - (12) depends on the bias voltage $V_G$, which affects the electron Fermi energy $\mu$. At moderate bias voltages, the ${\mathcal F}$ versus $V_G$ is given by Eq.~(3)
with 
$\mu \simeq \mu_D + \mu_0eV_G/(\mu_0 + \mu_D)$, where $\mu_0 = (\hbar^2v_W^2\kappa)/8e^2W$ with 
$v_W \simeq 10^8$~cm/s being the characteristic electron velocity in GCs. 
If $\kappa = 4 - 6$ and $W=10$
~nm, one obtains $\mu_0 \simeq (14 - 21)$~meV.
At low bias voltages ${\mathcal F} \simeq {\mathcal F}_0 = (\Delta_M/T)\exp(-\Delta_M/T)$. 
Considering the GC-FET Al/b-P/GC structures   with $\mu_D = 140$~meV and 
the Ti/b-As/GC structures
 with $\mu_D = 120$~meV,
for room temperature we obtain ${\mathcal F}_0 \simeq 0.113$ and ${\mathcal F}_0 \simeq 0.170$, 
respectively.

We estimate the electron thermal conductivity as $h \simeq v_W^2/2\nu$.
For $\nu = (1 - 2)$~ps$^{-1}$, the latter yields $h \simeq                                                                                                                                                                                                                                                                                                                                                                                                                                          (2.5-5.0)\times 10^3$~cm$^2$/s. These values are smaller than the record electron thermal conductivity, which can be extracted from the experimental papers~\cite{23,24} for the 2DEG 
in the suspended GCs with the electron densities $\Sigma \gtrsim 10^{13}$~cm$^{-1}$.
The lower values assumed by us can be justified by the presence of the gate layer and the substrate
surrounding the GC in the devices under consideration.
Thus, the thermal length (at a small $V_G$)                                                                                                                                                                                                                                                                                                                                                                                                                                                                                                                                                                                                                                                                                                                                                                                                                                                                                                                                                                                                                                                                                                                                                                                                                                                                                     

\begin{eqnarray}\label{eqA1}
{\mathcal L}_0 \simeq \displaystyle\sqrt{\frac{v_W^2\tau_{\varepsilon}}{2\nu(1 +\tau_{\varepsilon}{\mathcal F}_0\Delta_C/\tau_{\bot}T)}}
\end{eqnarray}
 depends on the quality of the GC and the interfaces between the GS and the surrounding layers (via the dependence on $\nu$) and on the contribution of the surface optical phonons to the electron energy relaxation time (via the dependence on $\tau_{\varepsilon}$).
This implies that the value of ${\mathcal L}_0$ can vary in a wide range.
For  definiteness, we  fix $\tau_{\varepsilon}$ and $\tau_{\bot}$, calculating the GC-FET detector characteristics for different $\nu$.

Setting 
  $\tau_{\varepsilon}\simeq 10$~ps,
$\tau_{\bot} = 10$~ps,  ${\mathcal F}_0 = 0.113 - 0.170$, and $\nu = (0.5 - 2.5)$~ps$^{-1}$,
we find ${\mathcal L}_0 \simeq (0.9 - 2.2)~\mu$m. 

The quantity ${\tilde \nu} \simeq \xi_{visc} q_P^2 \simeq \xi_{visc}(\pi/2L)^2
(\omega/\Omega_P)^2$, where $\xi_{visc} \simeq (250 -1000)$~cm$^2$/s~\cite{38}
is the electron viscosity coefficient and $q_{P} = (\pi\omega/2L\Omega_P)$ is the wavenumber of the  plasma mode  in the GC-FET under consideration. For the above values of $\xi_{visc}$ and $L = 1~\mu$m,
one obtains ${\tilde \nu} \simeq (0.06- 0.25)$~ps$^{-1}$.

\subsection*{Appendix B. Fermi energy vs gate voltage }
\setcounter{equation}{0}
\renewcommand{\theequation} {B\arabic{equation}}

When the electron plasma in the GC is degenerate ($\mu \gg T$) 
 $\mu$ is governed by the following equation:

 \begin{eqnarray}\label{eqB1}
 \mu= \hbar\,v_W\sqrt{\pi\Sigma_D + \frac{\kappa{\overline \Phi}}{4eW}},
  \end{eqnarray}  
 where the term proportional to ${\overline \Phi} = (\Delta_C - \Delta_M - \mu +eV_G)/e$ 
 corresponds to the electron density electrically induced in the GC (its dependence on $\mu$ reflects the quantum capacitance effect~\cite{53,54,55}).
Under the condition of the BL flat band ${\overline \Phi} = \Delta_C - \Delta_M - \mu_D =0$
 at $V_G =0$, hence $\mu =\mu_D$, and $\Sigma_D = \Sigma_D^{=} = (\Delta_C - \Delta_M)^2/\pi\hbar^2v_W^2 $. Considering this, we arrive at

 \begin{eqnarray}\label{eqB2}
 \mu= \sqrt{\mu_D^2+ \mu_0(eV_G +\mu_D -\mu )}
  \end{eqnarray} 
with $\mu_0= (\kappa\hbar^2v_W^2/8e^2W)$.

For  moderate and high gate voltages Eq.~(B2) yields 
  
 \begin{eqnarray}\label{eqB3}
\mu \simeq  \mu_D + \frac{ \mu_0}{(\mu_D+\mu_0)}eV_G,\, \mu \simeq  \sqrt{\mu_D^2 + 2\mu_0eV_G},
\end{eqnarray}
respectively.

In particular, using Eqs.~(2) and (B3), for the dc current-voltage characteristics in  the most actual range of  moderate  gate voltages, we obtain:

 \begin{eqnarray}\label{eqB4}
{\overline j} \simeq j^{max} \exp\biggl(-\frac{\Delta_M}{T_0}\biggr)\exp\biggl[\frac{\mu_0}{{(\mu_D + \mu_0)}}\frac{eV_G}{T_0}\biggr]\nonumber\\
\times\bigg[1 - \exp\biggl(-\frac{eV_G}{T_0}\biggr)\biggr].
\end{eqnarray}

\subsection*{Appendix C. Tunneling}
\setcounter{equation}{0}
\renewcommand{\theequation} {C\arabic{equation}}

At sufficiently high gate voltages when $e{\overline \Phi} > \Delta_C - \mu$, the energy barrier in the BL can become triangular, where $e{\overline \Phi} =\Delta_C - \Delta_M -\mu +eV_G = \mu_D- \mu +eV_G$ and
The latter inequality yields $eV_G > \Delta_M$.

In this case, the electron tunneling from the GC into the gate (via the triangular barrier in question) can be substantial.

 The tunneling electron current  through the triangular potential barrier assumed above [see Fig.~1(b)],
 can be estimated as ~\cite{56}

\begin{eqnarray}\label{eqC1} 
{\overline j}^{tunn} \simeq j^{max} \exp\biggl[- \frac{4\sqrt{2m}(\Delta_C -\mu)^{3/2}W}{3e\hbar\,{\overline \Phi}}\biggr],
\end{eqnarray}
 where 
 $m$ is the electron effective mass in the BL.

 Accounting for Eqs.~(2)and (B4), for the net DC current at large gate voltages we obtain 

\begin{eqnarray}\label{eqC2}
 \frac{{\overline j}}{j^{max}} \simeq \exp\biggl(\frac{\mu   - \Delta_C}{{\overline T}}\biggl)\nonumber\\
+ \exp\biggl[- \frac{4\sqrt{2m}(\Delta_C -\mu)^{3/2}W}{3e\hbar\,{\overline \Phi}}\biggr].
\end{eqnarray} 

Considering that a $e{\overline \Phi} = \mu_D - \mu + eV_G \simeq eV_G$, 
Eq.~(C2) becomes

\begin{eqnarray}\label{eqC3}
\frac{{\overline j}} {j^{max}} \simeq 
\exp\biggl(\frac{\mu   - \Delta_C}{{\overline T}}\biggl)\nonumber\\
+ \exp\biggl[- \frac{4\sqrt{2m}(\Delta_C -\mu)^{3/2}W}{3\hbar\,eV_G}\biggr]
.
\end{eqnarray} 
Hence,
the thermionic current  exceeds the tunneling current if
 
 \begin{eqnarray}\label{eqC4}
 V_G < \frac{4\sqrt{2m(\Delta_C - \mu)}T}{3e\hbar}W.
 \end{eqnarray} 
 Accounting for that the responsivity maximum corresponds to $\Delta_C-\mu = T$, we obtain
 
\begin{eqnarray}\label{eqC5}
 V_G  < V_G^{tunn} =\frac{4\sqrt{2m}T^{3/2}}{3e\hbar}W.
 \end{eqnarray}
 Assuming $m = 0.22m_0 =2\times 10^{-28}$~g and  $W = 10$~nm for room temperature we
 find $ V_G^{tunn} \simeq 0.14$~V.
 
 In  devices with relatively large $\Delta_M$ (in which even at $V_G =0$ the energy barrier in the BL is triangular, i.e., $e{\overline \Phi} > 0$ at $V_G =0$),                                                                                                                                                                                                                                                                                                                                                                                                                                                                                                                                                                                                                                                                                                                              the quantity $\overline \Phi$ can be small even
 at high bias voltages $V_G$, so that $V_G^{tunn}$ might be fairly large.

\end{document}